# Entanglement and the measurement problem

**Art Hobson**[1]

[1]Physics Professor Emeritus, University of Arkansas, Fayetteville, AR 72701, USA.  Email: ahobson@uark.edu

**Abstract**
The entangled "measurement state" (MS), predicted by von Neumann to arise during quantum measurement, seems to display paradoxical properties such as multiple macroscopic outcomes.  But analysis of interferometry experiments using entangled photon pairs shows that entangled states differ surprisingly from simple superposition states.  Based on standard quantum theory, this paper shows that (i) the MS does not represent multiple detector readings but instead represents nonparadoxical multiple *statistical correlations between* system states and detector readings, (ii) exactly one outcome actually occurs, and (iii) when one outcome occurs, the other possible outcomes simultaneously collapse nonlocally.  Point (iii) resolves an issue first raised in 1927 by Einstein who demonstrated that quantum theory requires instantaneous state collapse.  This conundrum's resolution requires nonlocal correlations, which from today's perspective implies the MS *must* be an entangled state.  Thus, contrary to previous presumed proofs of the measurement problem's insolubility, we find the MS to be the collapsed state and just what we expect upon measurement.

# 1 | INTRODUCTION

Physicists agree that Schrodinger's equation describes the evolution of non-relativistic quantum states between measurements, but there is no agreement on how states change during measurements. In fact, an apparent problem arises when one applies standard quantum theory (minus the collapse postulate) to measurements. John von Neumann analyzed the problem[1] and we follow his argument here. "Measurement" is the experimental determination and macroscopic recording of the value of a physical observable associated with a quantum system. von Neumann showed that, unless the system happens to be in an eigenstate of the measured observable, measurement leads to a "measurement state" (MS) whose mathematical representation is an entangled state that seems to predict the detector to be in a macroscopic superposition of exhibiting all of the possible outcomes, a paradox known as the "problem of outcomes."

This paper shows, based on standard quantum theory without a collapse postulate, that this is a pseudo-problem and that, far from predicting superposed outcomes, von Neumann's MS predicts an instantaneous collapse to a single eigenstate.[2] Specifically, we demonstrate, with no assumptions other than standard quantum physics (minus the collapse postulate): (i) The MS has been misinterpreted and does not in fact predict paradoxical multiple macroscopic *outcomes;* it instead correctly predicts non-paradoxical multiple *statistical correlations between* system and detector outcomes. (ii) Exactly one outcome actually occurs. (iii) The entanglement entails that, when one outcome occurs, the other outcomes simultaneously and nonlocally remain "dark" (i.e. do not occur). This resolves an objection to quantum physics first raised by Einstein in 1927.

That is, we show von Neumann's enigmatic MS to be in fact the collapsed state expected upon measurement. The collapse is derived (i.e. demonstrated) with no assumptions beyond the other standard principles of quantum physics.

We show the problem of outcomes arises from a mistaken understanding of entangled superpositions, not only in measurements but also in purely microscopic processes. The MS is obviously a superposition of subsystem product states. But the MS is poorly understood because no previous work has analyzed its complete phase dependence. Do the states of individual subsystems vary with phase, as they do in simple superpositions? If not, then what does vary with phase, i.e. precisely which entities occupy indefinite states when two or more subsystems are entangled? Such questions show that we do not fully understand the MS's phase dependence, i.e. we don't fully understand the MS.

We investigate these questions by studying earlier Bell-test quantum-optics experiments that measure momentum-entangled two-photon states. These experiments study a purely microscopic entangled superposition (mathematically identical with the MS) across all phases.  The results show that entangled superpositions differ sharply from simple superpositions, i.e. the implication of the "plus" sign differs surprisingly.  The MS does not represent multiple detector readings, but instead represents multiple *statistical correlations between* system states and detector readings.  These correlations are not paradoxical:  a macroscopic detector that simultaneously exhibited two states would be paradoxical, but a detector that simultaneously participates in two correlations is not paradoxical.  Furthermore, the MS directly implies that exactly one of these correlations is realized as the measurement outcome.  This resolves the so-called "Schrodinger's cat paradox," which in turn resolves the quantum measurement problem.[3]

The measurement problem has a long and rich history[3,4] that we will not try to comprehensively cover here.  The present analysis shares certain features with the modal interpretations of quantum physics.[5-8]  Like the present paper, modal interpretations are based on standard quantum mechanics but without the projection postulate (von Neumann's "Process 1");[1] they are realistic in the sense that they presume quantum systems possess real physical properties and they provide an objective (independent of humans) description of a single physical reality; they presume the dynamical state tells us the possible properties of the system and their corresponding probabilities; and they presume the dynamics (for non-relativistic systems) is governed by the unitary Schrodinger evolution and by the entanglement process presented in Section 2.  However the present analysis differs importantly from the modal interpretations' conclusion that "the dynamical state *never collapses* during its evolution."[5]  On the contrary, this paper deduces from quantum theory and from experimental evidence that an instantaneous nonlocal collapse takes place, resulting in one outcome occurring while the other outcomes simultaneously do not occur.  Thus, while the present paper does not *postulate* collapse, it *derives an instantaneous collapse* as a consequence of entanglement.

Other formulations also avoid postulating collapse.  Hugh Everett's many-worlds interpretation assumes "that there are many worlds which exist in parallel" and that a different branch is realized in each different world.[9,10]  David Bohm's hidden variables theory[11] assumes that a field (represented by the wave function) and particles are both present, with the field guiding the particles. Mario Bunge's realistic formulation[12] assumes a representative system/detector interaction and derives the Schrodinger evolution of the composite system, allowing one to deduce each observable's value from the detector reading. Gottfried and Yan[13,14] argue that, for all practical purposes, the off-diagonal terms of the exact density operator arising from the MS can be ignored, and that this solves the measurement problem.  A recent information-based analysis[15] regards the MS as the result of quantum measurement (as does this paper) and introduces, in addition to the measured quantum system and its measurement apparatus, a "programming system" that interacts with the quantum system and apparatus to encode the basis information for the system and apparatus, thus avoiding any classical concepts within an information-complete quantum formalism.

At least six papers--Lev Landau,[16] Gerhart Luders,[17] Josef Jauch,[18] Roland Omnes,[19] Stefan Rinner and Ernst Werner,[20] and S. Perez-Berrliaffa, G. Romero and H. Vucetich[21]--express the effect of measurement as a trace on the composite-system density operator representing the MS. This trace operation, which predicts definite outcomes at the subsystems, seems to yield just what we want, namely that measurement transforms the subsystem states into mixtures over the possible eigenstates. However, several objections are commonly raised against this proposal.[22] The present paper, in contrast, argues that the MS directly represents the collapsed state of the composite system, that the collapse is a consequence of entanglement, and that the entanglement is required in order to ensure a simultaneous (hence instantaneous) collapse over the separated branches.

Section 2 presents von Neumann's derivation of the MS, poses the measurement problem, reviews eight presumed measurement problem insolubility proofs, and explains why their conclusion cannot be correct.

Section 3 presents a crucial clue. In 1927, Einstein noted that quantum theory implies that measurements entail instantaneous collapse and suggested this would violate Special Relativity. However, today (unlike 1927) we know that instantaneous nonlocal changes of correlations violating Bell's inequalities really occur,[23-25] that they do not violate special relativity,[26] and that they occur when disparate systems are entangled.[27, 28] Thus from today's perspective, *Einstein's argument implies that nonlocal correlations are required during measurements*, entailing that entanglement is also required.

Is von Neumann's MS in fact precisely what we want? The answer can only come from fully understanding the MS, particularly its full phase dependence which has not, to this author's knowledge, been previously discussed in connection with the measurement problem. To this end, Section 4 reviews two 1990 quantum optics experiments exploring an entangled two-photon microscopic state that is mathematically identical to the MS.

The correct understanding of the microscopic version of the MS is then worked out in Section 5. We find that both entangled sub-systems are in definite (i.e. non-superposed) states, but that the *degree of correlation between these states is in an indefinite state*. This key new finding is summarized in Table 1.

Section 6 applies this new insight to the MS. We find the MS is not a macroscopic superposition of different detector *states*. It instead represents *different correlations between* detector states and system states. This is not paradoxical. Furthermore, quantum theory directly implies that precisely one outcome is realized. This resolves the Schrodinger's cat paradox.

Section 7 concludes the analysis by studying a particularly simple measurement example that also typifies the essentials of quantum measurements.

Section 8 summarizes the results and considers why it has taken so long to straighten out this simple misunderstanding of von Neumann's MS.

## 2  |  ENTANGLEMENT AND MEASUREMENT

The superposition postulate entails that, if $|A1\rangle$ and $|A2\rangle$ are Hilbert space vectors ("kets") representing possible states of a quantum system $A$, then all normalized linear superpositions of $|A1\rangle$ and $|A2\rangle$ also represent possible states of $A$. For example,

$$|\Psi_A\rangle = \frac{|A1\rangle + |A2\rangle}{\sqrt{2}} \qquad (1)$$

($|A1\rangle$ and $|A2\rangle$ orthonormal) represents a possible state of $A$. The superposition postulate is prerequisite to the Hilbert-space representation of quantum states, and the basis for conceptualizing quantum states as physically real waves in a quantum field that fills the universe.[26,29-31] $|\Psi_A\rangle$ represents a situation in which $A$ is represented neither by $|A1\rangle$ nor by $|A2\rangle$ but incorporates aspects of both, including "overlap" effects such as interference. As Dirac[32] put it, $A$ goes "partly into each of the two components" and "then interferes only with itself."

If quantum system $A$ interacts with another quantum system $B$, it frequently happens that the situation of $A$ and $B$ are then represented by an entangled superposition such as

$$|\Psi_{AB}\rangle = \frac{|A1\rangle|B1\rangle + |A2\rangle|B2\rangle}{\sqrt{2}}, \qquad (2)$$

where $|Ai\rangle$ and $|Bi\rangle$ ($i=1,2$) are orthonormal kets representing the "subsystems" $A$ and $B$, respectively. Although the physical interpretation of simple superpositions such as $|\Psi_A\rangle$ is clear, the physical interpretation of entangled superpositions such as $|\Psi_{AB}\rangle$ is not comparably clear. $|\Psi_{AB}\rangle$ is a superposition of two products $|Ai\rangle|Bi\rangle$ ($i=1,2$). $|A1\rangle|B1\rangle$ is commonly interpreted to represent a state of the composite system $AB$ in which $A$ has the properties associated with $|A1\rangle$ and $B$ has the properties associated with $|B1\rangle$, and similarly for $|A2\rangle|B2\rangle$. But if this is the case, then the physical interpretation of the state $|\Psi_{AB}\rangle$ would seem to be that $AB$ simultaneously exhibits properties associated with |A1> and |B1> AND properties associated with |A2> and |B2>, where "AND" represents the superposition. In the case of Schrodinger's iconic cat,[33] this would imply that the nucleus is both decayed and undecayed and, more disturbingly, the cat is both alive and dead. This paper will demonstrate that both quantum experiment and quantum theory show that this is actually *not* the case. Instead, the state $|\Psi_{AB}\rangle$ entails merely that $|A1\rangle$ and $|B1\rangle$ are *coherently* (in a phase-dependent manner) *correlated* with each other AND $|A2\rangle$ and $|B2\rangle$ are coherently correlated with each other (see Sections 4 and 5). This is not paradoxical.

Quantum measurements are important examples of entanglement. As first discussed by John von Neumann,[1,22] a "measurement" is the determination of the value of an observable associated with a quantum system *A*. If *A* happens to be represented by an eigenvector of the measured observable, then a good measurement will detect the associated eigenvalue. But what if *A* is represented by a superposition of eigenvectors of the measured observable? It will suffice for this paper's purpose to assume that *A*'s Hilbert space has only two dimensions, and that *A* is represented by the superposition Equation (1). The kets $|Ai\rangle$ *(i=1,2)* define the eigenvectors of the measured observable. We assume the existence of a detector *B* designed to distinguish between the $|Ai\rangle$.

For example, $|A1\rangle$ and $|A2\rangle$ could represent the paths of an electron passing through the slits of a double-slit apparatus, and *B* could be an electron detector for the "which-slit" observable whose eigenvectors are the $|Ai\rangle$. To make the which-slit measurement, *B* must distinguish between the states represented by $|A1\rangle$ and $|A2\rangle$, so *B* must contain a specific quantum detection component having quantum states represented by kets $|Bi\rangle$ such that, if *A* is in the state represented by $|Ai\rangle$, then detection yields the state represented by $|Bi\rangle$ (*i* = 1, 2).. Assuming a minimally-disturbing measurement that leaves eigenstates unaltered, and letting $|B_{ready}\rangle$ represent the state of *B*'s quantum component prior to measurement, the process

$$|Ai\rangle|B_{ready}\rangle \implies |Ai\rangle|Bi\rangle \quad (i=1,2) \qquad (3)$$

describes a measurement of the which-slit observable when *A*'s state is represented by an eigenstate. When *A* is in the state represented by $|\Psi_A\rangle$ and *B* measures the which-slit observable, simple linearity of the time evolution implies

$$\frac{(|A1\rangle + |A2\rangle)}{\sqrt{2}}|B_{ready}\rangle \implies \frac{|A1\rangle|B1\rangle + |A2\rangle|B2\rangle}{\sqrt{2}} = |\Psi_{AB}\rangle \qquad (4)$$

Thus von Neumann's straightforward argument shows the measurement creates the entangled superposition $|\Psi_{AB}\rangle$ of Equation (2), where "*B*" now refers to the quantum detection component of the detector.

But von Neumann's measurement postulate[1] implies that, when the which-slit observable is measured, *A* collapses into one of its eigenstates while *B* collapses into the corresponding detector state.[3,32] It is by no means clear that $|\Psi_{AB}\rangle$ (Equation (4)) represents such a measurement outcome. As Myrvdal[3] puts it, "The problem of what to make of this is called the 'measurement problem'." This paper will show that $|\Psi_{AB}\rangle$ does in fact represent the collapsed state and the single definite outcome expected from von Neumann's measurement postulate.

We have used the same notation, $|\Psi_{AB}\rangle$, for the arbitrary entangled state represented by Equation (2) (where $A$ and $B$ are arbitrary quantum systems) and for the specific case of the entangled state that develops when a detector measures a quantum system, represented by Equation (4) (where $B$ is now a detector). We will refer to this state in the context of Equation (4) as the "measurement state" (MS). We will also, however, need to refer to the arbitrary entangled state Equation (2), especially in Sections 4 and 5 where we analyze an experiment involving two microscopically entangled photons.

Thus the question of how to interpret entangled states looms large in the foundations of quantum physics. As noted above, the interpretation of general entangled states such as the one represented by Equation (2) is already murky as compared with the interpretation of simple superposition states such as the one represented by Equation (1). The problem of interpreting the MS is especially important, because macroscopically distinct states now lie on each side of the "plus" sign on the right-hand side of Equation (4). As already discussed, the superposition $|\Psi_A\rangle$ can be interpreted to represent a situation in which $A$ incorporates properties represented by both $|A1\rangle$ and $|A2\rangle$. And a product state such as $|A1\rangle|B1\rangle$ represents a state of the composite system $AB$ in which $A$ is represented by $|A1\rangle$ and $B$ is represented by $|B1\rangle$. Thus $|\Psi_{AB}\rangle$ appears to describe a detector that simultaneously "points" to two macroscopically different outcomes $|B1\rangle$ and $|B2\rangle$! The detector seems to display no definite outcome, a conundrum known as the "problem of outcomes".[3,22,33-47]

Such a superposition state would be paradoxical. Schrodinger, who imagined a cat attached to the detector in such a way that $|B1\rangle$ included a live cat and $|B2\rangle$ included a dead cat, described $|\Psi_{AB}\rangle$ as representing a "living and dead cat …smeared out in equal parts."[33] As one quantum foundations expert writes,

> The crucial difficulty is now that it is not at all obvious how one is to regard the dynamical evolution described by [Equation (4)] as representing measurement in the usual sense. This is so because [Equation (4)] is …not sufficient to directly conclude that the measurement has actually been completed.[22]

In fact, while measurement should lead to a specific eigenstate of the measured observable, Equation (4) appears to entail that "the system has been sucked into a vortex of entanglement and no longer has its own quantum state. On top of that, the entangled state fails to indicate any particular measurement outcome."[48]

As noted, it seems paradoxical that quantum measurements lead to a state represented by the MS. Measurement should lead to a situation in which $A$ is represented by one of its eigenvectors $|Ai>$ and $B$ is represented by the corresponding $|Bi>$. Since quantum uncertainty typically implies unpredictable measurement outcomes, it is reasonable to conclude that measurement should lead to a state represented by an ignorance-interpretable mixture[22] of the products $|A1>|B1>$ and $|A2>|B2>$. Assuming the initial state is represented by $|\Psi_A>$, such a post-measurement mixture would be represented by the density operator

$$\rho_{mixed} = (|A1>|B1><B1|<A1| + |A2>|B2><B2|<A2|) / 2 \qquad (5)$$

This mixture can be interpreted as "the system is represented by a single component $|Ai>|Bi>$, but we cannot know whether $i=1$ or $2$ until we look at the outcome."

Beginning with von Neumann's analysis, at least eight "measurement problem insolubility proofs"[1,49-55] have assumed that, in order to obtain definite outcomes, the measurement process should transform the composite system $AB$ into a mixture analogous to Equation (5). The initial state of $A$ is assumed to be pure and to be represented by a superposition such as Equation (1). The analysis then investigates whether a suitable composite-system post-measurement mixture can be reached via a unitary process. To achieve this, the detector must be represented by a mixture initially, because unitary processes cannot turn a pure state into a mixture. Since $B$ is macroscopic, such an initial mixture seems appropriate. Thus von Neuman and seven succeeding analysts asked: Is there an initial mixed-state density operator $\rho_{ready}$ of $B$ and a unitary process $U$ acting on $AB$ such that $U$ transforms the initial composite density operator $|\Psi_A><\Psi_A|\otimes\rho_{ready}$ into the desired composite mixture? The eight insolubility proofs showed, with varying assumptions, the answer is "no," presumably demonstrating the measurement problem to be insoluble.

Section 3 will show, however, that the premise of these insolubility proofs, namely that Equation (5) represents the appropriate post-measurement state, was doomed from the start, precisely because it is *not* entangled and thus cannot have the properties required if quantum theory is to describe the measurement process. To put this another way, there are reasons why the post-measurement state *must* be an entangled state, which implies that it *cannot* be a mixture such as Equation (5). Sections 4-7 then show that the MS does have the desired properties.

## 3 | A CRUCIAL CLUE FROM EINSTEIN

At the 1927 Solvay Conference, five years prior to von Neumann's analysis[1] of quantum measurement, Einstein asked the audience to consider an experiment in which electrons pass through a tiny hole in an opaque screen and then impact a large hemispherical detection screen centered at the hole (Fig. 1).

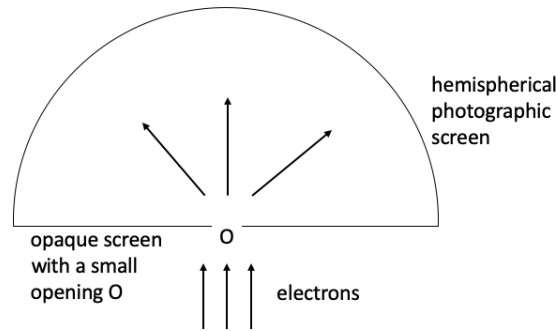

**Figure 1.** Einstein's thought experiment. Each electron diffracts widely, reaching the entire screen simultaneously, yet only one point shows an impact. How do the other points *instantaneously* remain dark? Does this violate Special Relativity?

According to the Schrodinger equation, each electron's state diffracts widely, spreading and reaching the entire screen simultaneously. Yet each electron registers at only a single point. How, Einstein asked, do the other points *instantaneously* remain "dark," i.e. not show an impact? As Einstein put it in his notes, this "entirely peculiar mechanism of action-at-a-distance, which prevents the wave continuously distributed in space from producing an effect in *two* places on the screen," presents a fundamental problem. It appears to imply instant signaling, violating special relativity.[56-58]

Einstein's argument shows that, under a realistic and objective (independent of humans) interpretation of quantum physics, the Schrodinger equation is at odds with experimental facts unless the electron's state collapses instantaneously and nonlocally upon measurement.[26,29] Thus, realistic quantum physics implies *instantaneously-established nonlocal correlations are essential to the measurement process*. Indeed, Fuwa et al.[59] experimentally verified the nonlocal character of the measurement transition. Since nonlocality is essential to measurement, the presumed post-measurement mixed state was doomed from the start precisely because it does *not* exhibit the required nonlocality. But entangled superpositions do exhibit the required nonlocality.[27,28] So from our modern point of view, Einstein's argument shows that entanglement, far from being an unwelcome paradox, is *required* in measurements. This is a crucial clue and good news for quantum foundations, because von Neumann's predicted MS is just such an entangled superposition!

Gottfried and Yan's argument[13,14] should be mentioned. Their resolution of the measurement problem "diagonalizes the density operator." They form the exact density operator $\rho = |\Psi_{AB}\rangle \langle \Psi_{AB}|$, which can be written

$$|\Psi_{AB}\rangle \langle \Psi_{AB}| = \rho_{diagonal} + \rho_{off\text{-}diagonal} \qquad (6)$$

where $\rho_{diagonal} = \rho_{mixed}$ (Equation (5)) and

$$\rho_{off\text{-}diagonal} = (|A1\rangle|B1\rangle \langle B2|\langle A2| + |A2\rangle|B2\rangle \langle B1|\langle A1|) / 2. \qquad (7)$$

Recall that the exact expectation value of any observable $F$ is

$$\langle F \rangle = Tr(\rho F) = \Sigma_j \Sigma_k \; \rho_{jk} F_{kj} \qquad (8)$$

where $\rho_{jk}$ and $F_{kj}$ are matrix elements of $\rho$ and $F$. Gottfried and Yan argue that off-diagonal terms in Equation (8) can be ignored because they involve matrix elements such as $\langle B1|\langle A1|F|A2\rangle|B2\rangle$ that are non-zero only for a "fantastic" observable $F$ because $|B1\rangle$ and $|B2\rangle$ represent the states of widely separated detectors. In Gottfried and Yan's opinion, matrix elements for such fantastic observables can, for all practical purposes, be neglected so that we can replace $\rho$ by $\rho_{mixed}$. But we have seen that this premise is doomed because $\rho_{mixed}$ lacks the required nonlocal properties, so Gottfried and Yan's proposal fails.

## 4 | EXPERIMENTAL STUDIES OF STATES HAVING ENTANGLED SPATIAL PATHS

Sections 2 and 3 presented the measurement problem and some previous research on the problem. Sections 4-7 will present a suggested resolution. This Section reviews interferometry experiments and theory that investigate the *microscopic* entangled superposition $|\Psi_{AB}\rangle$ Equation (2) over its full 0-to-π range of phases, for a system of two momentum-entangled (i.e. path-entangled) photons. The results provide a key insight into solving the measurement puzzle.

As preparation, we first study the simple superposition Equation (1). Consider the interferometer experiment of Figure 2. On each experimental trial, a photon enters a 50-50 beam splitter *BS1* which transforms the photon's state into the superposition Equation (1) where $|A1\rangle$ and $|A2\rangle$ respectively represent paths 1 and 2. A series of single-photon trials probes this state using mirrors *M* that bring the two branches together, phase shifters $\phi_1$ and $\phi_2$ that lengthen the two paths by phases $\phi_1$ and $\phi_2$, and a second beam splitter *BS2* that mixes the branches together. Measurement occurs at photon detectors *B1* and *B2*.

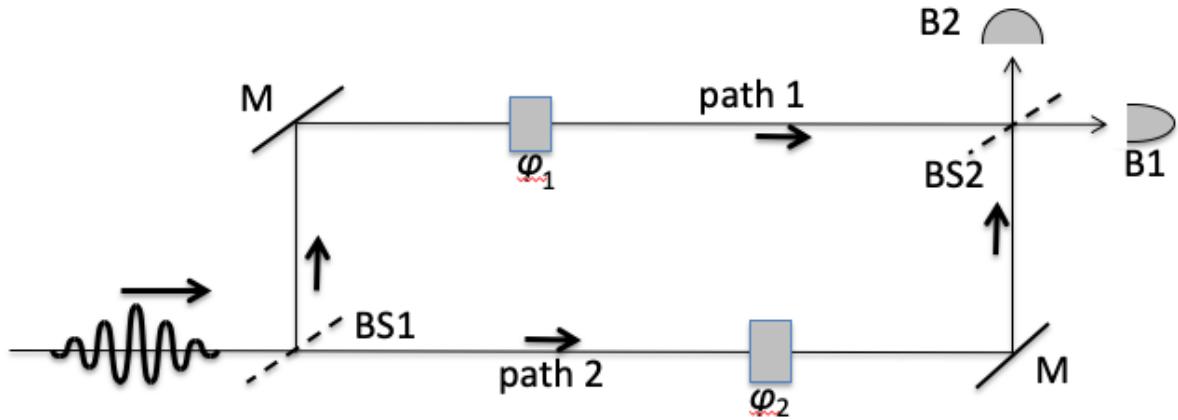

**Figure 2.** Mach-Zehnder interferometer experiment. A photon traverses a beam splitter, travels on two phase-shifted paths to another beam splitter, and is detected.

Figure 3 shows the results. Varying $\phi_1$ through 180 degrees causes the photon's state to shift from 100% probability of detection at *B1*, through diminishing probabilities at *B1* and increasing probabilities at *B2*, finally reaching 100% probability of detection at *B2*. The photon exhibits similar interference upon varying $\phi_2$. Note that *A*'s state depends only on the phase *difference* $\phi_2 - \phi_1$.

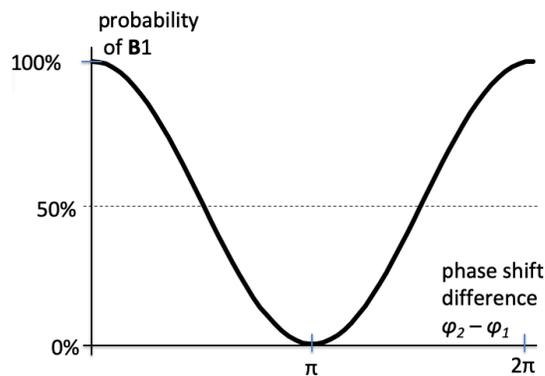

**Figure 3.** Single photon interference, pointing to Dirac's conclusion that "each photon ...interferes only with itself," i.e. each photon follows both paths.

Since single-trial results vary regardless of which phase shifter varies, it is hard to avoid the conclusion that each photon follows both paths. In fact, let us assume the contrary, namely that each photon follows only one path. Suppose the phase shifters are set to ensure 100% probability of detection at *B1*. Under our assumption, this setting guarantees that every photon following path 1, and every photon following path 2, is detected at *B1*. Suppose path 2 is now blocked without changing the phase settings, so that (still under our one-path assumption) every detected photon must now follow path 1 and be detected at *B1*. But the experiment shows that, to the contrary, 50% of the detected photons now go to *B2*. Conclusion: each photon follows both paths. For a full discussion, see.[26,29]

As Paul Dirac put it, "The new theory, which connects the wave function with probabilities for one photon, gets over the difficulty by making each photon go partly into each of the two components. Each photon then interferes only with itself."[32] This illustrates why the "plus" sign in a superposition such as Equation (1) is interpreted by the word "and."

We turn now to the entangled state Equation (2). Many Bell inequality tests beginning with Clauser's[60] and Aspect's[61] used polarization-entangled photon pairs to study the full phase dependence of this state. More useful for this paper are interferometer experiments by Rarity and Tapster[62] and Ou, Zou, Wang and Mandel[63-65] conducted nearly simultaneously in 1990. Both of these "RTO experiments" (for Rarity, Tapster, and Ou et al.) used *momentum*-entangled photon pairs to conduct Bell inequality tests of the entangled superposition Equation (2).

Figure 4 shows the layout. The source creates entangled pairs of photons $A$ (moving leftward) and $B$ (moving rightward) by laser down-conversion in a non-linear crystal. The down-converted photons are prepared in the entirely microscopic state represented by $|\Psi_{AB}\rangle$ (Equation (2)) by selecting four single-photon beams, each a plane wave having a distinct momentum (i.e. wave vector), from the output of the crystal. Figure 4 resembles two back-to-back Mach-Zehnder interferometer experiments (Figure 2) with BS1 located effectively inside the source.

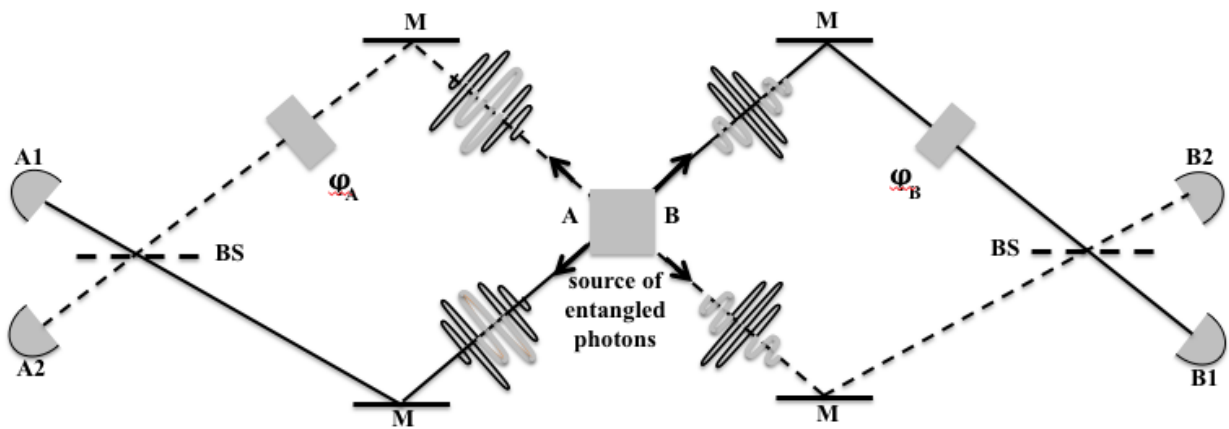

**Figure 4.** The RTO experiments. In each trial, each of two entangled photons travels two phase-shifted paths (one solid, the other dashed) to beam splitters and coincidence detectors. Think of one biphoton spreading from the source along both the solid and the dashed paths.

For simplicity and clarity, Figure 4 differs from the layout shown in RTO's reports. In Figure 4, paired photons are directed oppositely. This arrangement would result if the entanglement were prepared by the cascade decay of an atom as in Reference.[66] In RTO's experiments, however, down-converted photon pairs are emitted into two angular cones, resulting in photons that are not oppositely directed. Figure 4's simpler geometry is pedagogically useful and has no effect on our arguments.

Although Figure 4 represents each photon as a wave packet spreading along two paths directed leftward or rightward, the composite system *AB* should be regarded as a *single* object, a "biphoton." In each trial, a biphoton spreads outward from the source along two superposed branches. One branch, represented by the first term |*A1*>|*B1*> in Equation (2), spreads along the solid path and the other branch, |*A2*>|*B2*>, spreads along the dashed path. As the biphoton *AB* moves outward along the solid path, *A* encounters a mirror *M*, then a beam splitter *BS* where it transmits and reflects to detectors *A1* and *A2*; photon *B* encounters a mirror *M*, a phase shifter $\phi_B$, and a beam splitter *BS* where it transmits/reflects to detectors *B1*/*B2*. The other half of the entanglement, namely the dashed path, has a similar description. The experiments record outcomes at four photon detectors equipped with coincidence timers.

Horne, Shimony and Zeilinger[66,67] predict RTO's results theoretically and we follow their optical-path analysis here. They begin by calculating the two-point nonlocal quantum field amplitudes *Ψ(Ai, Bj)* at the four coincidence detectors *(Ai, Bj)*, and from these results they predict single-photon results. For example, *Ψ(A1, B2)* has two contributions, one from phase shifts in the beam following the solid path (the first term in |*Ψ<sub>AB</sub>*>) and the other from the dashed path (the second term). From Equation (2), assuming distinct plane waves exp(i**k**•**x**) for each single-photon beam,

$$\Psi(A1,B2) = \{exp(i\phi_w)exp[i(\phi_x+\phi_B)] + exp[i(\phi_y+\phi_A)]exp(i\phi_z)\}/2\sqrt{2} \quad (9)$$

where $\phi_w$, $\phi_x$, $\phi_y$, $\phi_z$ are fixed phase-shifts resulting from mirrors and beam splitters, and the additional factor of 1/2 comes from the superpositions created at the two beam splitters. Using the Born rule, Equation (9) implies the coincidence probability

$$P(A1, B2) = |\Psi(A1,B2)|^2 = [1+cos(\phi_B-\phi_A + \phi_v)]/4 \quad (10)$$

where $\phi_v$ is a fixed phase arising from $\phi_w$, $\phi_x$, $\phi_y$, $\phi_z$. Similarly,

$$P(A1, B1) = [1+cos(\phi_B-\phi_A + \phi_u)]/4, \quad (11)$$

where $\phi_u$ is another fixed phase. Remarkably, the sinusoidal terms predict coherent (phase-dependent) nonlocal interference between *A* and *B*, regardless of their separation. There are similar expressions for *P(A2, B1)* and *P(A2, B2)*.

Single-photon predictions then follow. For example, from simple probability theory

$$P(A1) = P(A1, B1) + P(A1, B2)$$
$$= [1+\cos(\phi_B-\phi_A + \phi_u)]/4 + [1+\cos(\phi_B-\phi_A + \phi_v)]/4. \quad (12)$$

Horne et al. then show the two fixed phase factors $\phi_u$ and $\phi_v$ differ by $\pi$:

$$\phi_v = \phi_u + \pi \ (mod\ 2\pi). \quad (13)$$

Thus the sinusoidal terms in Equation (12) *interfere destructively*, and *$P(A1) = 1/2$ regardless of phase*. Equations (12) and (13) show this remarkable result to arise from destructive interference of two phase-dependent nonlocal contributions from the distant *other* photon B! The result at all four single-photon detectors is the same:

$$P(A1) = P(A2) = P(B1) = P(B2) = 1/2. \quad (14)$$

Unlike the non-entangled single-photon superposition $|\Psi_A\rangle$, where the superposed photon is coherent (phase-dependent) as shown by Figure 3, each entangled photon is "decohered"[22] and cannot interfere with itself. Instead the two photons interfere with each other despite being separated by an arbitrary distance. More accurately, *each biphoton interferes with itself*. Thus no single-photon interference fringes are associated with the state represented by $|\Psi_{AB}\rangle$.

Special relativity entails that this must be the case: Since single-photon phase dependence could be used to establish an instantaneous communication channel between *A* and *B*, entanglement *must* deprive individual photons of their phase. The result is nonlocal coherence of the biphoton, and decoherence of individual photons. Decoherence is required by special relativity.

Equation (14) can also be derived by tracing the pure state density operator $|\Psi_{AB}\rangle\langle\Psi_{AB}|$ over one subsystem to obtain the density operator for the other subsystem.[22] This yields two density operators that appear to be mixtures but are not really "ignorance mixtures" as the word "mixture" is usually understood because the biphoton is in fact not in a mixed state but rather in a pure state represented by $|\Psi_{AB}\rangle$. The optical path analysis, above, derives Equation (14) while avoiding these controversial[22] subtleties.

A few definitions can put these predictions into more comprehensible form: If one photon is detected in state 1 and the other in state 2, the two outcomes are said to be "different." Otherwise, the outcomes are the "same." Then from Equations (10) and (11), and similar expressions for *P(A2, B1)* and *P(A2, B2)*,

$$P(same) = P(A1,B1) + P(A2,B2) = 1/2[1 + \cos(\phi_B - \phi_A)] \qquad (15)$$

$$P(different) = P(A1,B2) + P(A2,B1) = 1/2[1 - \cos(\phi_B - \phi_A)]. \qquad (16)$$

Their difference, graphed in Figure 5, is called the "degree of correlation":

$$C = P(same) - P(different) = \cos(\phi_B - \phi_A). \qquad (17)$$

Section 5 explores its physical significance.

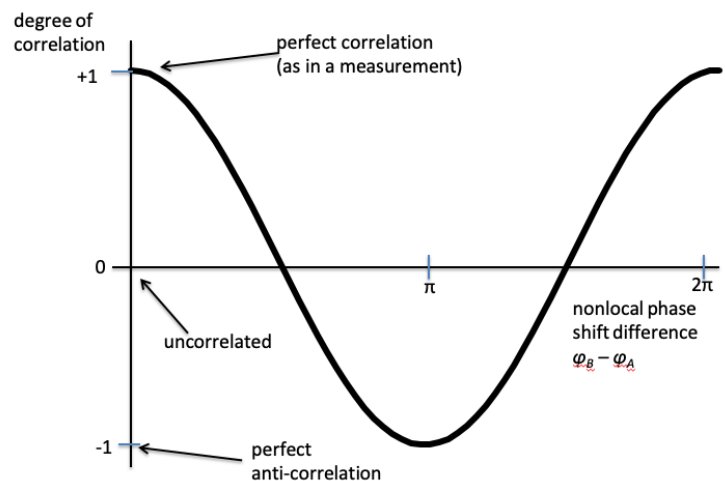

**Figure 5.** Results of the RTO experiment, demonstrating nonlocality (violation of Bell's inequality). The two photons interfere with each other across an arbitrary distance, i.e. each biphoton interferes with itself.

## 5 | INTERPRETATION OF ENTANGLED MICROSCOPIC STATES

The original purpose of RTO's experiments was to demonstrate violations of Bell's inequality by comparing theoretical predictions, Figure 5, with experimental measurements. The experimental results agreed with Figure 5 and violated Bell's inequality by 10 standard deviations, confirming the nonlocal nature of $|\Psi_{AB}\rangle$.

What does Figure 5 mean conceptually? At zero phase difference, where the two phase shifters are set to equal phases, P(same) = 1 and C = P(same) - P(different) = +1. Thus both stations always agree, despite the presence of beam splitters that randomize each photon prior to detection (see Figure 4). It is as though coins were flipped at each station and they always came out either both heads or both tails! Zero is the "measurement" phase angle where B's state is perfectly (and instantaneously[23-25]) correlated with A's state. The nonlocality is intuitively obvious: Each photon acts like a detector of the state of the other photon regardless of separation! Simply based on this conclusion, we can make an important observation about the entangled MS Equation (4): *Nonlocality is a central feature of quantum measurements*.

For small non-zero phase differences, there is a small probability P(different) that results at the two stations will differ, i.e. observation of *B* no longer provides reliable information about *A*. With increasing phase difference, this unreliability increases until, at π/2, the two detector pairs are entirely uncorrelated, and C=0. As the phase further increases from π/2 to π, P(different) increases while P(same) decreases, making C more and more negative. Finally, C = -1 at phase difference π, implying perfect anti-correlation. Thus *C* is aptly called the "degree of correlation."

This description gives us a clear sense of the physical meaning of the fully entangled state Equation (2), indicating precisely which entities are superposed. The biphoton's phase controls the degree to which the fixed phase-independent 50-50 states of its two spatially separated subsystems are *statistically correlated*. Compare this with the phase of the simple superposition $|\Psi_A\rangle$ (Equation (1)), which controls the degree to which the single system *A* is represented by one or the other *state*. The entities before and after the plus signs in Equations (1) and (2) are conceptually quite different: Equation (1) sums two *states* while Equation (2) sums two *correlations between states*.

This distinction is crucial. A *state* is a situation (or configuration or path) of a single quantum object, but a *correlation* is a statistical relationship between two (or more) quantum objects. A *superposition* is the simultaneous existence of two or more states of a single quantum object. An *entanglement* is the simultaneous existence of two or more relationships (specifically, correlations) between the states of two or more quantum objects. Creating an entanglement is quite a different matter from creating a superposition.

To elaborate, Table 1 compares the superposition represented by $|\Psi_A\rangle$ (columns 1-2) with the entanglement represented by $|\Psi_{AB}\rangle$ (columns 3-5) at five different phases. Column 2 demonstrates interference between the states represented by $|A1\rangle$ and $|A2\rangle$, implying the photon is in a superposition of following both paths and that the *state of A* varies with phase. The phase dependence in column 2 shows that *A* interferes with itself.

In contrast, column 4 shows that, when the subsystems are represented by the pure state $|\Psi_{AB}\rangle$, neither photon has a phase. Thus neither photon can interfere with itself, so neither photon can be represented by a superposition state. They are decohered. Both photons are represented by fixed, phase-independent, 50-50 states at all phase angles, just as though they were in ignorance mixtures (which they are not). But phase dependence has not vanished, it has only been transferred to the composite system. As column 5 reveals, the *degree of correlation between the fixed states of A and B* now varies with phase.

| Simple superposition of 1 photon. | | Entangled superposition of 2 photons | | |
|---|---|---|---|---|
| $\phi_2-\phi_1$ | State of photon | $\phi_B-\phi_A$ | State of each photon | Correlation between photons |
| 0 | 100% 1, 0% 2 | 0 | 50-50 1 or 2 | 100% corr, 0% anticorr |
| $\pi/4$ | 71% 1, 29% 2 | $\pi/4$ | 50-50 1 or 2 | 71% corr, 29% anticorr |
| $\pi/2$ | 50% 1, 50% 2 | $\pi/2$ | 50-50 1 or 2 | 50% corr, 50% anticorr |
| $3\pi/4$ | 29% 1, 71% 2 | $3\pi/4$ | 50-50 1 or 2 | 29% corr, 71% anticorr |
| $\pi$ | 0% 1, 100% 2 | $\pi$ | 50-50 1 or 2 | 0% corr, 100% anticorr |

**Table 1.** Comparison between a simple superposition (Fig, 2) and an entangled superposition (Fig. 4). In Fig. 2, single-photon states vary with phase. In Fig. 4, only the *correlation between single-photon states* varies with phase while single-photon states have no phase. Thus *each biphoton is coherent* but its subsystems are incoherent. That is, entanglement *decoheres* each photon while transferring coherence to the biphoton.

A photon represented by $|\Psi_A\rangle$ is in a coherent (phase-dependent) superposition of being in two states (i.e. of following two *paths*). $|\Psi_{AB}\rangle$, on the other hand, represents the coherent superposition of two *correlations between fixed states*. Instead of two coherent *states* existing simultaneously, two coherent *relationship between states* exist simultaneously. *Neither subsystem is "smeared"* (as Schrodinger apparently believed); instead, *only the relationship between subsystems is smeared*. Briefly, $|\Psi_A\rangle$ is a superposition of *states* and $|\Psi_{AB}\rangle$ is a superposition of *correlations*.

Thus $|\Psi_{AB}\rangle$ is qualitatively different from $|\Psi_A\rangle$. $|\Psi_A\rangle$ exhibits properties of $|A1\rangle$ AND $|A2\rangle$, where "AND" indicates the superposition. If you amplify $A$ to macroscopic dimensions, you will get a macroscopic superposition. $|\Psi_{AB}\rangle$ exhibits properties of correlations between $|A1\rangle$ and $|B1\rangle$ AND correlations between $|A2\rangle$ and $|B2\rangle$. If you amplify $A$ and $B$ to macroscopic dimensions, you will *not* get a macroscopic superposition, you will simply get correlations between macroscopic objects. The entanglement process transfers the coherence (phase dependence) of each photon to correlations between the two photons, leaving individual photons in mixtures that are incoherent but that are not ignorance mixtures. $|\Psi_{AB}\rangle$ is a "superposition of correlations between properties," in contrast to $|\Psi_A\rangle$ which is a "superposition of properties."

There is a better way to think about all this: Regard $AB$ as a single object, a biphoton. Then Equation (2) describes a superposition of this object. In the RTO experiment (Figure 4), the two superposed states are represented by the solid line and the dashed line. In this context, it makes no sense to speak of the superposition of a single subsystem, but it does make sense to speak of the superposition of the biphoton. It is the biphoton that goes through the phases graphed in Figure 5 and indicated in Table 1 column 5. Both branches (solid and dashed) of the biphoton exist simultaneously.

## 6 | INTERPRETATION OF THE MEASUREMENT STATE

Section 4 analyzed the microscopic state represented by Equation (2) mathematically, and Section 5 interpreted this state physically. We now apply these insights to the entangled MS of a quantum system *A* and its detector *B* as derived in Equation (4).

In order for *B* to be a reliable detector, its states must be perfectly correlated with *A*'s states--it must exhibit $|Bi\rangle$ when and only when *A* is represented by $|Ai\rangle$ (*i = 1, 2*). Thus Figure 5 implies the MS must be established at zero non-local phase: $\phi_B - \phi_A = 0$. At this phase, two nonlocal perfect statistical correlations between a phase-independent 50-50 state of *A* and the corresponding phase-independent 50-50 state of *B* exist simultaneously. As shown in Section 5, *contrary to Schrodinger's description,*[33] *neither subsystem state can be "smeared out" (superposed) because neither subsystem has a phase.* Instead, *correlations between* fixed states of *A* and *B* are smeared as shown in Table 1, while the detector indicates a single definite outcome.

Applying Table 1 to Schrodinger's example,[33] the cat is predicted to be alive in 50% of trials, dead in the other 50%, and never in both states simultaneously. Phase alterations would not smear the cat, they would smear only the correlations between the cat and the nucleus leading not to a smeared cat but only to imperfect detection. There is no paradoxical macroscopic superposition.

But if neither *A* nor *B* is superposed, what *is* superposed? What does the MS's "plus" sign really mean? The answer, from Table 1 at zero phase: $|A1\rangle$ is perfectly correlated with $|B1\rangle$ AND $|A2\rangle$ is perfectly correlated with $|B2\rangle$, where "AND" represents the superposition. This simply says *both correlations exist simultaneously:* $|A1\rangle$ if and only if $|B1\rangle$ AND $|A2\rangle$ if and only if $|B2\rangle$. Again, there is no paradoxical macroscopic superposition. It's only the *correlations (relationships) between states,* not the states themselves, that are superposed.

Entanglement transforms a superposition of 2 *states* into a superposition of *two correlations between states.* This makes quantum measurements possible because subsystem states can then be amplified to macroscopic dimensions without requiring the creation of a macroscopic superposition. Neither subsystem is in a macroscopic superposition.

Since neither subsystem is superposed, only a single outcome occurs--a conclusion that also follows from Equation (14). This single definite outcome occurs instantly upon entanglement, as facilitated by the nonlocal properties of the entangled MS.[23-25] Thus we have derived the collapse as an inevitable consequence of entanglement, and have no need to postulate such a process. The MS is the collapsed state. Our conclusion follows merely from standard principles of quantum theory with no other assumptions.

So von Neumann's enigmatic measurement state, Equation (4), is just what we want. This entangled pure state provides the desired correlations, a single outcome, and the nonlocality required by Einstein's argument. Note that the collapse is established at the microscopic level, prior to macroscopic amplification. The next Section provides an example of the sequence of events.

## 7 | EXAMPLE

The following simple example typifies quantum measurements and illustrates the preceding insights in terms of a specific measurement process.

Consider the set-up in Figure 6. A single photon traverses a beam splitter, creating the superposition represented by Equation (1) whose branches correspond to separate paths toward widely separated photon detectors. Analogously to Figure 1, we assume the two detectors are equidistant from the beam splitter. Each detector contains a photo-sensitive plate that, upon absorbing a photon, releases an electron.

Von Neumann's argument implies that, as the two branches of the superposition approach the detectors, at some point the branches overlap the detectors sufficiently that the entanglement process represented by Equation (4) occurs, where $|B_{ready}\rangle$ denotes the microscopic state of the detectors prior to entanglement while $|B1\rangle$ and $|B2\rangle$ denote their states following entanglement but prior to amplification and macroscopic recording.

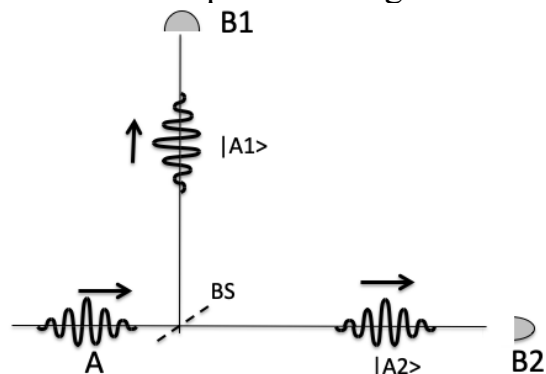

Figure 6. A simple measurement that mirrors Einstein's example (Fig. 1): A photon passes through a beam splitter and is measured by photon detectors. Due to entanglement between the two superposed photon branches and the detectors, one electron in one detector absorbs the photon's energy while the other detector simultaneously remains "dark."

At the instant of interaction, the state jumps from a superposition of two paths of $A$ (Equation (1)) to a superposition of two correlations between $A$ and $B$ (right-hand side of the process in Equation (4)). This entangled state is not paradoxical. The right-hand side of Equation (4) entails precisely the proper correlations: $|A1\rangle$ if and only if $|B1\rangle$, AND $|A2\rangle$ if and only if $|B2\rangle$. The excitation is transferred to only *one* detector while the other detector remains unexcited. More correctly, either the solid branch or the dashed branch of the superposed biphoton (Figure 4) is randomly selected. In fact, Fuwa et al.[59] show experimentally and theoretically that the set-up shown in Figure 6 leads to entanglement and that the predicted nonlocal collapse occurs; the nonlocality of the collapse is verified quantitatively by the experimental violation of an EPR-steering inequality.

Thus the collapse, a non-linear and irreversible process, occurs at the microscopic level. Once one photoelectron is released, the process is thermodynamically irreversible because the electron is released into a vast number--a continuum--of free electron states and cannot feasibly be reversed. This electron triggers an avalanche of other electrons leading to a macroscopic mark at one detector.

Other measurement set-ups follow the same general principles. For example, in the measurement described by Einstein (Section 3), each small region of the detection screen acts as a single detector, and the diffracted electron's quantum state entangles with all these many regions. Thus the argument above, involving only two detectors, must be extended to N detectors.

## 8 | SUMMARY AND CONCLUSION

Using only the standard principles of quantum physics, but minus the collapse postulate, we have shown that quantum state collapse occurs as a consequence of the entanglement that occurs upon measurement as described in 1932 by von Neumann (Equation (4)). The entangled "measurement state" of a quantum system and its detector is the collapsed state: It incorporates the required perfect correlations between the system and its detector, it predicts precisely one definite outcome, and it incorporates the nonlocal properties--the instantaneous collapse across all branches of the superposition--that Einstein showed to be required in quantum measurements.

The measurement state Equation (4) does not describe a detector in a paradoxical superposition of displaying multiple outcomes, as had been supposed by Schrodinger and others. Instead, quantum theory concludes that this state entails just what we expect following a measurement: The states represented by $|A1\rangle$ and $|B1\rangle$ are perfectly correlated, AND the states represented by $|A2\rangle$ and $|B2\rangle$ are perfectly correlated, where "AND" represents the plus sign in the mathematical representation of the state. Entanglement entails merely the simultaneous occurrence of two correlations between subsystems, not the simultaneous occurrence of two individual states of either subsystem. There can be no paradoxical superposition of different detector states or of different system states, because the entanglement has shorn both the detector and the quantum system of their quantum phases.  The phase has been transferred from the individual subsystems to the degree of correlation between subsystems.

To put all of this more directly, the single quantum object *AB* (the biphoton) collapses from a superposition to one of its members.

The measurement state's entanglement and its nonlocal properties, far from being paradoxical, are required in order to guarantee that the collapse occurs simultaneously across all branches of the superposition.  Eight previous insolubility proofs failed because they did not incorporate this required nonlocality.  Nonlocality is a central feature of quantum measurement.

There is no need for a special collapse postulate because the entangled state is the collapsed state.  Collapse occurs instantly upon entanglement.

This analysis should not be regarded as one more interpretation of quantum physics.  It is instead a correction of the previous misunderstanding of von Neumann's entangled measurement state.  It is not surprising that this misunderstanding has persisted for nearly 90 years.  After all, entanglement and nonlocality are deeply involved in the measurement problem's proper resolution but they only began to be understood in 1964,[69] leading to a long period of gradual acceptance with confirmation only in 2015.[23-25]  The delay in understanding measurement stemmed from this delay in understanding nonlocality.

**ACKNOWLEDGEMENTS**
I thank University of Arkansas colleague Suren Singh for many consultations, and for showing me that a biphoton really is a single thing.  I thank University of Arkansas colleagues Julio Gea-Banacloche, Michael Lieber, Peter Milonni, and Barry Ward.  I thank Nathan Argaman, Kenneth Augustyn, Rodney Brooks, Mario A. Bunge, Zeng-Bing Chen, Klaus Colanero, Allan Din, Moses Fayngold, Ron Garret, Kurt Gottfried, David Green, Robert Griffiths, Lucien Hardy, Ulrich Harms, Norbert Ibold, Ken Krechmer, Andrew Laidlaw, Franck Laloe, Gui-Lu Long, James Malley, Tim Maudlin, N. David Mermin, Peter Morgan, Michael Nauenberg, Oliver Passon, Jakub Ratajczak, Ravi Rau, Stefan Rinner, Gustavo E. Romero, Jacob van Dijk, and Herve' Zwirn.

Art Hobson        Entanglement and the measurement problem                     2326. Hobson A. Realist analysis of six controversial quantum issues. Matthews M (ed). *Mario Bunge: A Centenary Festschrift.* Springer; 2019:329-348.
27. Gisin N. Bell's inequality holds for all non-product states. *Phys Lett.* 1991;154:201-202.
28. Liang Y, Masanes L, Rosset D. All entangled states display some hidden nonlocality. *Phys. Rev. A*;2012;86:052115.
29. Hobson A. There are no particles, there are only fields. *Am J Phys.* 2013;81:211-223.
30. Long G. Realistic interpretation of quantum mechanics and encounter-delayed-choice experiment. *Science China: Phys, Mech & Astron.* 2018;61:03031.
31. Long G. Collapse-in and collapse-out in partial measurement in quantum mechanics and its wise interpretation. *Science China: Phys, Mech & Astron.* 2021;64:280321.
32. Dirac P. *The Principles of Quantum Mechanics.* Clarendon Press; 1947.
33. Schrodinger E. The present situation in quantum mechanics: a translation. *Proc Am Phil Soc.* 1937;124:323-338.
34. van Kampen N. Ten theorems about quantum mechanical measurements. *Physica A.* 1988;153:97-113.
35. Bell J. Against measurement. *Phys World* 1990;3:33-40.
36. Ney A. Introduction. In Ney A, Albert D. *The Wave Function.* Oxford University Press; 2013:1-51.
37. Allori V. Primitive ontology and the structure of fundamental physical theories. In Ney A, Albert D., *ibid.* 58-75.
38. Wallace D. A Prolegomenon to the ontology of the Everett interpretation. In Ney A, Albert D. *ibid.* 203-222.
39. Sherrer R. *Quantum Mechanics: An Introduction.* Pearson; 2006.
40. Ballentine L. *Quantum Mechanics: A Modern Development.* World Scientific; 1998.
41. Griffiths D. *Introduction to Quantum Mechanics.* Cambridge Press; 2018.
42. Greiner W. *Quantenmechanik. I, Einfuehrung.* Springer;1994.
43. Rae A. *Quantum Mechanics.* Institute of Physics; 2002.
44. Ghirardi G, Rimini A, Weber T. Unified dynamics for microscopic and macroscopic systems. *Phys. Rev. D.* 1986;34:470-491.
45. Maudlin T. Three measurement problems. *Topoi.* 1995;14:7-15.
46. Leggett A. Quantum measurement problem. *Science.* 2005;307:871-872.
47. Leggett A. Schrodinger's cat and her laboratory cousins. *Contemporary Phys.* 2009;50:243-258.
48. Schlosshauer M. *Elegance and Enigma.* Springer; 2011:141-142.
49. Wigner E. The problem of measurement. *Am. J. Phys.* 1963;31:6-15.